# Efficient Parallel Computation of the Estimated Covariance Matrix

ODED GREEN, Georgia Institute of Technology, Technion
LIOR DAVID, Technion
AMI GALPERIN, Technion
YITZHAK BIRK, Technion

Computation of a signal's estimated covariance matrix is an important building block in signal processing, e.g., for spectral estimation. Each matrix element is a sum of products of elements in the input matrix taken over a sliding window. Any given product contributes to multiple output elements, thereby complicating parallelization. We present a novel algorithm that attains very high parallelism without repeating multiplications or requiring inter-core synchronization. Key to this is the assignment to each core of distinct diagonal segments of the output matrix, selected such that no multiplications need to be repeated yet only one core writes to any given output-matrix element, and exploitation of a shared memory (including L1 cache) that obviates the need for a corresponding awkward partitioning of the memory among cores. Implementation on Plurality's HyperCore shared-memory many-core architecture demonstrates linear speedup of up to 64 cores and speedups of ~85$X$ for 128 cores. On an x86 system we demonstrate that the new algorithm has consider parallel speedups but also show that a sequential implementation of the new algorithm outperforms the parallel implementation of the baseline approach. On a quad-core x86 system, the new algorithm is 20$X$ faster than sequential baseline and 5$X$ than parallel implementation of the baseline.

Categories and Subject Descriptors:

**B.3.2 [Design Styles]** : Shared memory**, D.1.3 [Concurrent Programming]** : Parallel programming**, F.2.1 [Numerical Algorithms and Problems]**Computations on matrices

## 1. INTRODUCTION

Covariance estimation is widely used for signal processing and even for cryptanalysis. Sliding window averaging (aka ``sub-aperture averaging'' and ``the covariance method'') is currently a prominent autocorrelation (and covariance) estimator[Kay 1988; Marple 1987].

Computation of the estimated covariance matrix essentially entails the averaging of inner products within a sliding window over the input matrix: for each window position, a vector is formed by column-stacking the columns of the window, and is then multiplied by its conjugate transpose. Averaging those results over all possible positions of the window within the input matrix yields the estimated covariance matrix.

The product of any two input-matrix elements usually contributes to multiple output-matrix elements, yet matrices may be large and the number of distinct two-element products may be huge, so efficient computation on a real system poses apparently conflicting challenges: 1) computational efficiency (specifically, avoiding repetition of multiplications) and 2) efficient use of memory (both reducing the memory footprint and improving locality, the latter for efficient use of caches). For parallel implementations, additional challenges include 3) partitioning the work into many pieces and 4) prevention of contention and synchronization requirements among compute cores.

In this paper we present a novel, efficient and highly parallel way to compute the estimated covariance matrix, jointly addressing all the aforementioned challenges.

(The results remain unchanged.) Key to our success is a unique partitioning of the output matrix elements among compute tasks, combined with the use of a shared memory (including level-1) many-core architecture, such as Plurality's HyperCore [Plurality 2007; Plurality 2009], in order to circumvent the resulting awkward memory partitioning. This partitioning is also relevant to x86 systems.

Our focus has been on parallelization. Nonetheless, our approach also improves the efficiency of sequential implementations due to reduction in computation requirements, as will be explained later.

The remainder of this paper is organized as follows. Section 2 formulates the problem, Section 3 describes the new parallel algorithm, Section 4 presents the detailed derivations and proofs, Section 5 provides experimental results, and section 6 presents concluding remarks.

## 2. THE COVARIANCE METHOD
### 2.1 Background

The use of estimated covariance matrices originated from the area of speech processing [Makhoul 1975]. One way to compute the estimated covariance matrix is by the Covariance Method. The method's main rationale is minimizing the error in the estimate of a covariance matrix of a time series. Good estimates can give insights pertaining to the data periodicities, and enable fast and accurate analysis of the input data.

The covariance method is a biased estimator, and its output is a Hermitian, positive semi-definite matrix, so it is guaranteed to be non-singular. This is unlike the output of other methods, such as autocorrelation, and often causes the covariance method to be preferred [Marple 1987].

The method can be used for signal processing algorithms, e.g., whenever the correlation between signal history domain data samples is needed (as in[DeGraaf 1998]), synthetic aperture radar range-azimuth focusing[Lopez-Dekker and Mallorqui 2010] , to smooth spatial clutter by averaging over given transposed data(see [Fante et al. 1994]), or to perform 2D spectral analysis (e.g., [Jakobsson et al. 2000]). In all the aforementioned cases, the computation of the estimated covariance matrix is an important computational building block.

### 2.2 Formal expression of the estimated covariance matrix

*Terminology and Symbols*

All indices (rows and columns) start at 1, i.e., the first element in $\vec{V}$ is $\vec{V}(1)$, which is also denoted $\vec{V}_1$. The (NXM) input matrix is denoted A. $S^{p,q}$ denotes the sliding $P \times Q$ window (a block within A), where the $(p, q)$ superscripts denote the position of its upper left corner. The sliding window is also known as a sub-aperture. $\vec{V}^{p,q}$ is the column stack of $S^{p,q}$ such that

$$S^{p,q}(i, j) \equiv \vec{V}^{p,q}(P \cdot (j-1) + i). \tag{2.1}$$

The conjugate transposes of matrices and vectors are denoted by $A^H$ and $\vec{V}^H$, respectively, and C denotes the output matrix. For clarity of exposition, we will refer to elements of the input matrix A, while those of the output matrix C will be referred to as indices. The reader is referred to [Yadin et al. 2008] for additional reading and visualization of the window creation. Yadin et al. refer to the windows as sub-apertures.

The required result is

$$C = \sum_{p=1}^{N-P} \sum_{q=1}^{M-Q} C^{p,q} = \sum_{p=1}^{N-P} \sum_{q=1}^{M-Q} \vec{V}^{p,q} \cdot (\vec{V}^{p,q})^H,$$

(2.2)

and this expression can also be regarded as a serial algorithm for computing C. Note that C is Hermitian, being the sum of Hermitian matrices, so only the upper triangle or the lower triangle needs to be computed.

Also, there is a tradeoff between increasing the size of the sub-aperture is this increases the computational complexity and reduces the sampling space. Decreasing the size of the sub-aperture increases the sampling space at the expense of granularity.

## 2.3 Related Work

In [DeGraaf 1998]), the covariance matrix is used as part of the Minimum Variance Method (MVM) for Synthetic Aperture Radar (SAR) image processing. Due to the increase complexity of computing the estimated covariance method for MVM, DeGraaf suggested an alternative and computationally cheaper estimated covariance matrix as part of the Reduced-Rank MVM (RRMVM), however, this approach reduces the performance of the algorithm as it introduces more noise.

In [DeGraaf 1998]) an image of size $1600x1600$ is used by the MVM algorithm. The image is divided into $100x100$ apertures, which are the input matrices for the covariance method, of $25x25$ with an overlap between them. The sub-apertures are of size $10x10$. Yadin et al. use images of size $8000x8000$ with the MVM algorithm, this increases the total number of estimated covariance matrices computed to $250K$ matrices ($25X$ more matrices need to computed for these size images than that used by DeGraaf). Both papers show that MVM gives a better output image than using an FFT. Further, given the massive number of times the estimated covariance matrix is computed, reducing the computation time for each matrix is beneficial. Additional algorithmic parameters can be found in these papers.

In [Jakobsson, Marple and Stoica 2000] the covariance methods is compared to the Toeplitz-Block-Toeplitz method as part of the classic Capon estimator [Capon 1969] and APES (Amplitude and Phase) spectral estimator [Jian and Stoica 1996; Liu et al. 1998] and is shown to be more computationally demanding by about an order of magnitude.

We will show that the covariance method becomes more applicable due to a reduced computational complexity and faster sequential and parallel execution times.

## 2.4 Deficiencies of a Straightforward Serial Implementation

The multiplication of $\vec{V}^{p,q}$ by $(\vec{V}^{p,q})^H$ in (2.2) is actually a multiplication of every element in each vector by every element in the other vector. Therefore, when considering two sliding window positions $S^{p,q}$ and $S^{p',q'}$ such that both contain the product of some two elements x and y, both result matrices $C^{p,q}$ and $C^{p',q'}$ will contain the following products: $x \cdot \bar{x}, x \cdot \bar{y}, y \cdot \bar{x}, y \cdot \bar{y}$. The positions of these products relative to the upper left corner of the window will not be the same in the two matrices, as will the indices to which the product contributes. The result is repetition of these multiplications.

In Fig. 1, for example, the product $A_{3,3} \cdot \overline{A_{4,4}}$ is needed for every window that contains those two elements, two of these windows are presented. It would be desirable to compute it once and then write it to the correct place for each of the windows. (The challenge is to do so without consuming much memory for temporary results or requiring inter-task synchronization.)

### 2.5 Parallelization Challenges

Parallelization of the serial algorithm presents several challenges: 1) avoiding redundant multiplications, 2) obviating the need for synchronization and atomic instructions, 3) limiting the required amount of memory for intermediate results, 4) utilizing all cores all the time (efficient massive parallelism with load balancing), 5) fast dispatching of tasks to the cores, and 6) creating a "good" memory access pattern. These must all be addressed concurrently!

Various intuitive parallelization approaches fail to meet all challenges. For example, assigning a different row of the output matrix to each task would result in redundant multiplications. Simplistically assigning any given product to a single task would result in multiple tasks contributing to the same index of the output matrix, thus requiring synchronization and possibly atomic operations.

Another failed approach is one whereby each concurrent thread keeps a temporary copy $C^{p,q}$ of the estimated covariance matrix. A great deal of parallelism can be achieved in the multiplication stage, but in the subsequent summation stage most of the cores will not be utilized. Also, this approach requires sizable memory, larger than most caches, and the resulting cache misses would hurt performance. Finally, this approach does not avoid redundant multiplications.

We next present our novel parallel algorithm, which jointly addresses all the aforementioned challenges.

### 3. OUR PARALLEL ALGORITHM

### 3.1 Multiplication Combinations

In the previous section, it was shown that the product of any given pair of matrix elements may be required in several window positions. Also, those elements' positions relative to each other are the same regardless of the window. However, the product of any element pair in the context of each of the windows containing it will be written to a different index in the output matrix based on the position of the elements in the window (which is responsible for the creation of $\vec{V}^{p,q}$ ). We next proceed to show how efficient parallelization can be achieved despite this complexity.

*Definition*

| A1,1 | A1,2 | A1,3 | A1,4 | A1,5 | ... | A1,M |
|------|------|------|------|------|-----|------|
| A2,1 | A2,2 | A2,3 | A2,4 | A2,5 | ... | A2,M |
| A3,1 | A3,2 | A3,3 | A3,4 | A3,5 | ... | A3,M |
| A4,1 | A4,2 | A4,3 | A4,4 | A4,5 | ... | A4,M |
| A5,1 | A5,2 | A5,3 | A5,4 | A5,5 | ... | A5,M |
| ... | ... | ... | ... | ... | ... | ... |
| AN,1 | AN,2 | AN,3 | AN,4 | AN,5 | ... | AN,M |

Fig. 1. Uses of a product. The elements $A_{3,3}$ and $A_{4,4}$, and consequently $A_{3,3} \cdot \overline{A_{4,4}}$ , are contained in several sliding window positions. Two of these windows are presented, the gray window and the blue window. The middle section is the overlapping of these windows.

Let $A_{r_1,c_1}$ and $A_{r_2,c_2}$ be two input-matrix elements. Their inter-element distance (vector) is defined as:

$$(\Delta r, \Delta c) = (r_2 - r_1, c_2 - c_1). \tag{3.1}$$

*Definition*

A combination is the set of all products of two input-matrix elements with the same inter-element distance; it is denoted by this distance, $\Delta$, which must satisfy:

$$-(P-1) \leq \Delta r \leq P-1 \wedge -(Q-1) \leq \Delta c \leq Q-1. \tag{3.2}$$

The restrictions on the distances reflect the fact that we are only interested in products of elements that can be within the same PxQ window.

Note that combinations are based on the relative position of the elements in the input matrix. For example, the products of element pairs $(A_{3,3}, A_{4,4})$ and $(A_{5,5}, A_{6,6})$ both belong to combination $(\Delta r, \Delta c) = (1,1)$.

## 3.2 Unique Combinations

Recall that the estimated covariance matrix is Hermitian, so it suffices to compute the upper triangle or lower triangle sub-matrices. Also, any two inverse-distance combinations (e.g. $(\Delta r, \Delta c) = (a, b)$ and $(\Delta r, \Delta c) = (-a, -b)$) consist of the same element pairs. As the multiplication is done by taking the conjugate of one of the elements, we can utilize the trivial identity $A_{x',y'} \cdot \overline{A_{x,y}} \equiv \overline{A_{x,y} \cdot \overline{A_{x',y'}}}$ to further reduce the number of actual multiplications. Consequently, the products belonging to combinations that "own" the lower triangle needn't be computed.

The set of unique combinations is a union of two groups:

all combinations wherein the first multiplier is placed at $S^{1,1}$. There are $P \cdot Q$ such combinations. In this case $0 \leq \Delta r \leq P - 1$ and $0 \leq \Delta c \leq Q - 1$;

all combinations wherein the first element is in the first column and the second element is to the right and above the first element. Thus, for each row that the first element is in, it is possible to place the second element in $Q - 1$ different places. This allows for a total of $(P - 1) \cdot (Q - 1)$ different combinations. In this case: $-(P - 1) \leq \Delta r \leq -1$ and $1 \leq \Delta c \leq Q - 1$.

Accordingly, the set of unique combinations, denoted UC, is specified by modifying the distance restrictions of (3.2) as follows:

$$UC = \begin{Bmatrix} 0 \leq \Delta r \leq P-1 \\ 0 \leq \Delta c \leq Q-1 \end{Bmatrix} \cup \begin{Bmatrix} -(P-1) \leq \Delta r \leq -1 \\ 1 \leq \Delta c \leq Q-1 \end{Bmatrix}. \tag{3.3}$$

The total number of combinations is:

$$|UC| = P \cdot Q + (P-1) \cdot (Q-1). \tag{3.4}$$

As can be seen, $\Delta c$ is always positive, so the second element of any product is to the right of the first element. $\Delta r$, in contrast, can be positive or negative, so there is no keen observation to make on its value. A consequence of $\Delta c$ being positive is that all results are written to the upper triangle.

### 3.3 The Parallel Algorithm

The parallel algorithm offers a combination-centric solution rather than a window centric one: work is partitioned among tasks at combination granularity.

*Proposition*: each product is computed exactly once, so the number of multiplications is optimal.    (Obvious)    ∎

**The number of multiplications**

We now proceed to determine the number of unique products for a specific combination as follows. Place the combination such the leftmost product is on the left border of the matrix A and the uppermost product is on the top border of the matrix A. This ensures that both elements are in the input matrix and belong to at least one window, $S^{1,1}$. Next, select the two elements that are offset by one position to the right. The product of these two elements will be required by window $S^{1,2}$ and may also be required by $S^{1,1}$. It is possible to continue "stepping" the choice of both the elements to the right as long the right-most element does not exceed matrix borders. This can be done $(M - \Delta c)$ times. Afterwards, return to the left most position and take one step downward (both data elements). Repeat previous stage until the bottom-most element in a pair reaches the bottom border; this can be done $(N - \Delta r)$. For each combination, the number of unique products is thus given by:

$$\mu(\Delta r, \Delta c) = (N - \Delta r) \cdot (M - \Delta c). \tag{3.5}$$

Let UM denote the number of unique multiplications (products). The total number of multiplications that the parallel algorithm computes is thus:

$$UM = \sum_{\Delta r, \Delta c} \mu(\Delta r, \Delta c). \tag{3.6}$$

The computation of this sum can be found in Section 4.1, in which it is also compared with the number of multiplications that the straightforward serial algorithm carries out.

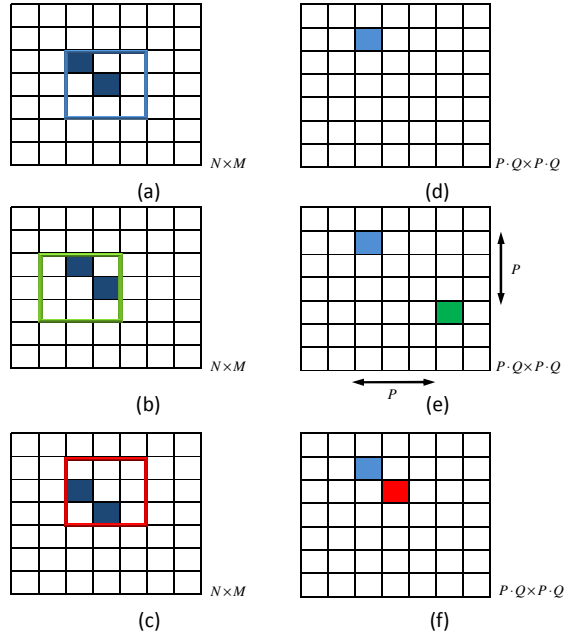

Fig. 2. The sliding window over an element pair. (a) Initial placement of the sliding window in the input matrix, surrounding the two elements; (b) moved one step leftward; (c) moved from its original position one step upward; (d) The output-matrix index to which the product is written in the context of the (a) window position. This index is also presented in (e) and (f); (e) the index to which the same product is written in the (b) positioning context; (f) the index to which the same product is written in the (c) positioning context.

**Write pattern (affected indices) of a combination**

*Definition*

Given two elements in the input matrix, $A_{r_1,c_1}$ and $A_{r_2,c_2}$, the initial placement of the sliding window for those two elements is its lowest and rightmost location such that the window contains the two elements and the two elements are on the left and top borders of the window. For convenience, we denote the initial placement by the location of its upper left corner.

The initial write indices for the product of element $(r_1, c_1)$ and $(r_2, c_2)$ are:

$$(r,c) \triangleq (r_1 + c_1 \cdot P, r_2 + c_2 \cdot P),$$
(3.7)

The sliding window may only be moved from its initial position Fig. 2 (a) to the left and upward as is depicted in Fig. 2 (b) and (c), respectively. Assuming that a product is written to $C_{r,c}$ when the window is in its initial position Fig. 2 (d), two important observations are made: 1) sliding the window one step to the left (b) results in writing the product to $C_{r+P,c+P}$, (see Fig. 2 (e)). 2) Sliding the window one step upward (c) results in writing the product to $C_{r+1,c+1}$ (see Fig. 2 (f)). In Fig. 2 (e) and (f) the write index of the first window is given in addition to the new write indices.

Based on the above, there are several additional important observations: 1) the target indices of any given combination lie on a diagonal segment (in the output matrix); 2) for elements located near the boundaries of A, it may not be possible to move the sliding window as it may exit the boundaries of A ; accordingly, some multiplications write to fewer result indices; 3) the total number of legal positions of the sliding window about a given multiplication (matrix-element pair) is up to $(P - |\Delta r|) \cdot (Q - |\Delta c|)$. We next prove this.

Given the two elements and their initial placement, it is possible to move the sliding window around them leftward and upward, as depicted in Fig. 2 (b) and (c). The maximum number of times that a sliding window can be moved to the left while staying within the bounds of the matrix A is $(P - |\Delta r|)$. Next, the sliding window is returned to its initial position Fig. 2 (a) and is then moved upward by one position. Next, it is again possible to move the sliding window a total of $(P - |\Delta r|)$ steps to the left. This is repeated a total of $(Q - |\Delta c|)$ times, so there are a total of $(P - |\Delta r|) \cdot (Q - |\Delta c|)$ legal positions for any given ``element-pair centric'' sliding window. This is the number of indices whose values are affected by a given product.

*Definition*

Let $\eta(\Delta r, \Delta c)$ be the function that expresses the maximum number of indices to which the combination $(\Delta r, \Delta c)$ writes:

$$\eta(\Delta r, \Delta c) = (P - |\Delta r|) \cdot (Q - |\Delta c|).$$
(3.8)

An implication of these observations is that the number of consecutive indices on a given diagonal that will be written to by a given combination depends on the number of times that the window can be moved upward, while the number of times that the window can be moved leftward will determine the number of disjoint segments that a given combination writes to on the given diagonal. For certain multiplications (mainly those that are near the borders of A), the result will not be written to all of the combination's designated indices.

The pseudo code for the new parallel algorithm can be seen in Algorithm 1.

---

**Algorithm 1.** New Parallel Algorithm

**parallel for on all combination ($\Delta r, \Delta c$):**

---

```
    foreach unique multiplication m in the combination:
        val ← A(r_1, c_1) · \overline{A(r_2, c_2)}.
        compute initial position of m.
        foreach valid shift, (r̂, ĉ), of m:
            C_{r̂,ĉ} ← C_{r̂,ĉ} + val
        end
    end
end
```

*Proposition:*
The target indices of the combinations jointly cover the entire upper triangle of the output matrix, and any given index is written to by a single combination. Consequently, no synchronization is required among the cores.

*Proof:*
In Section 4.2, we prove that different combinations do not write to (access) the same indices. In Section 4.3, we prove that the total number of indices that all the combinations write to is equal to the number of indices in the upper triangle. This shows that the all indices in the upper triangle are written to and that no synchronization is required to avoid concurrent writing of data to the same index.

### 3.4 Parallel Algorithm Efficiency

We have already shown that the number of multiplications is optimal. As each combination writes to a distinct set of output-matrix indices, as depicted in Fig. 3, it is possible to compute the different combinations concurrently on multiple cores and update the output indices with their contributions without allocating temporary matrices. For a shared-memory architecture in which the cache size is limited, this is very important, and there is no need for locks, synchronization or even atomic instructions.

Partitioning by combination thus requires the bare minimum number of multiplications, a dramatic reduction relative to the naive approach, and the work can be carried out concurrently on different cores with no need for synchronization. Whenever a core becomes available, a combination may simply be allocated to it. Thus, parallelism at the combination granularity is very effective, provided that the memory can be partitioned "by combination" among the cores is an effective manner.

## 4. FORMAL DERIVATIONS AND PROOFS

### 4.1 Complexity comparisons and approximations

We begin by presenting a computational complexity analysis of the serial algorithm. Next, we present the complexity analysis of the new algorithm.

*Naïve serial algorithm*
Computing each $C^{p,q}$ entails multiplying a $PQ \times 1$ matrix by a $1 \times PQ$ matrix. This reqires a total of $P^2Q^2$ multiplications. As each of the $C^{p,q}$ is added to C, an equal number of additions is required for each window. The number of $C^{p,q}$s depends on the number of possible positions of the sliding window. The window can be moved a total of $(N - P)$ steps downwards and a total of $(M - Q)$ steps to the right (for each of the downward movements). This makes a total of $(N - P) \cdot (M - Q)$ positions. We denote the number of multiplications required by the serial algorithm as SM and the number

of additions as SA. The number of multiplications used by the naïve serial algorithm is:

$$SM = (N-P)(M-Q)P^2Q^2. \tag{4.1}$$

The number of addition operations used by the naïve serial algorithm:

$$SA = (N-P)(M-Q)P^2Q^2. \tag{4.2}$$

*Our algorithm*

We have shown that by using the combination approach it is possible to compute a minimum number of multiplications and that number of multiplications that each combination executes is $\mu(\Delta r, \Delta c) = (N - \Delta r) \cdot (M - \Delta c)$. In order to compute the total number of multiplications, $\mu(\Delta r, \Delta c)$ needs to be computed for every combination. We denote the number of unique multiplications by UM. UM is computed as follows:

$$UM = \sum_{(\Delta r, \Delta c) \in UC} \mu(\Delta r, \Delta c) = UM_1 + UM_2 \tag{4.3}$$

$$UM_1 = \sum_{\Delta r=-(P-1)}^{0} \sum_{\Delta c=0}^{Q-1} (N-\Delta r)(M-\Delta c)$$

$$= \sum_{\Delta r=-(P-1)}^{0} (N-\Delta r) \sum_{\Delta c=0}^{Q-1} (M-\Delta c) = \frac{P}{2}(2N-P+1) \cdot \frac{Q}{2}(2M-Q+1) \tag{4.4}$$

$UM_2$ is computed in a similar fashion:

$$UM_2 = \sum_{\Delta r=1}^{P-1} \sum_{\Delta c=1}^{Q-1} (N-\Delta r)(M-\Delta c) = \frac{P-1}{2}(2N-P) \cdot \frac{Q-1}{2}(2M-Q) \tag{4.5}$$

Following this, UM is written as follows:

$$UM = \frac{P}{2}(2N-(P-1)) \cdot \frac{Q}{2}(2M-(Q-1)) + \frac{P-1}{2}(2N-P) \cdot \frac{Q-1}{2}(2M-Q) \tag{4.6}$$

In order to make the difference more readily visible, we approximate UM conservatively (we increase it). In $UM_2$, we will change $(P − 1)/2$ to $(P/2)$, and $(2N − P)$ to $(2N + 1 − P)$. The same changes will be made to expressions involving Q. We rewrite UM as follows:

$$UM \le \widehat{UM} = 2\frac{P}{2}(2N-(P-1)) \cdot \frac{Q}{2}(2M-(Q-1)). \tag{4.7}$$

The ratio of the number of multiplications executed by the two algorithms is:

$$\frac{SM}{UM} \ge \frac{SM}{\widehat{UM}} = \frac{(N-P) \cdot (M-Q) \cdot P \cdot Q}{2(N-(P-1)/2) \cdot (M-(Q-1)/2)} \tag{4.8}$$

Clearly, the numerator is always greater than the denominator. For N=M=32 and P=Q=13 ($P = 0.4M$ as is suggested by [Yadin, Olmar, Oron and Nathansohn 2008]), $SM/\widehat{UM} = 45.125$. \hfill (4.9)

### 4.2 Collision Freedom Among Combinations

*Theorem*
The write indices of two different combinations do not intersect.

*Proof*

Consider two different combinations, $(\Delta r_1, \Delta c_1)$ and $(\Delta r_2, \Delta c_2)$, such that

$$\begin{cases} \Delta r_1 = r_1 - r_2 \\ \Delta c_1 = c_1 - c_2 \end{cases} ; \begin{cases} \Delta r_2 = r_3 - r_4 \\ \Delta c_2 = c_3 - c_4 \end{cases},$$

(4.10)

and let e=Δr1-Δr2 and f=Δc1-Δc2.

Since the combinations are different, there are 3 scenarios:
1. $e \neq 0 \wedge f \neq 0$.
2. $e = 0 \wedge f \neq 0$.
3. $e \neq 0 \wedge f = 0$.

Consider the first scenario. Assume by contradiction that $e \neq 0 \wedge f \neq 0$, yet the two combinations write to the same index in the result matrix. From (4.10) and the definition of e, f:

$$\begin{cases} r_1 - r_3 = r_2 - r_4 + e \\ c_3 - c_1 = c_4 - c_2 - f \end{cases}.$$

(4.11)

As the two combinations (are assumed to) write to the same index, it follows from (3.7) that

$$(r_1 + c_1 \cdot P, r_2 + c_2 \cdot P) = (r_3 + c_3 \cdot P, r_4 + c_4 \cdot P).$$

(4.12)

Rewriting this vector in equation form:

$$\begin{cases} r_1 + c_1 \cdot P = r_3 + c_3 \cdot P \\ r_2 + c_2 \cdot P = r_4 + c_4 \cdot P \end{cases}.$$

(4.13)

Rearranging terms:

$$\begin{cases} r_1 - r_3 = P \cdot (c_3 - c_1) \\ r_2 - r_4 = P \cdot (c_4 - c_2) \end{cases}.$$

(4.14)

Carrying out a division operation for each equation yields the following result:

$$\begin{cases} P = \dfrac{r_1 - r_3}{c_3 - c_1} \\ P = \dfrac{r_2 - r_4}{c_4 - c_2} \end{cases}$$

(4.15)

Using these equations we get the following:

$$\dfrac{r_1 - r_3}{c_3 - c_1} = \dfrac{r_2 - r_4}{c_4 - c_2}.$$

(4.16)

Replacing the elements on the left hand side of the equation using (4.11) yields:

$$\dfrac{r_2 - r_4 + e}{c_4 - c_2 - f} = \dfrac{r_2 - r_4}{c_4 - c_2}$$

(4.17)

|       |       |       |       |       |       |     |        |
|-------|-------|-------|-------|-------|-------|-----|--------|
| C₁,₁  | C₁,₂  | C₁,₃  | C₁,₄  | C₁,₅  | C₁,₆  | ... | C₁,PQ  |
| C₂,₁  | C₂,₂  | C₂,₃  | C₂,₄  | C₂,₅  | C₂,₆  | ... | C₂,PQ  |
| C₃,₁  | C₃,₂  | C₃,₃  | C₃,₄  | C₃,₅  | C₃,₆  | ... | C₃,PQ  |
| C₄,₁  | C₄,₂  | C₄,₃  | C₄,₄  | C₄,₅  | C₄,₆  | ... | C₄,PQ  |
| C₅,₁  | C₅,₂  | C₅,₃  | C₅,₄  | C₅,₅  | C₅,₆  | ... | C₅,PQ  |
| C₆,₁  | C₆,₂  | C₆,₃  | C₆,₄  | C₆,₅  | C₆,₆  | ... | C₆,PQ  |
| ...   | ...   | ...   | ...   | ...   | ...   | ... | ...    |
| C_{PQ,1} | C_{PQ,2} | C_{PQ,3} | C_{PQ,4} | C_{PQ,5} | C_{PQ,6} | ... | C_{PQ,PQ} |

Fig. 3. Output Matrix. Indices of different colors are written to by different combinations.

Consequently, $e = 0 \wedge f = 0$, in contradiction with the assumption.

Using the same arithmetic manipulation, it is obvious that the 2nd and 3rd scenarios cannot happen either. Therefore, the two different combinations do not write to the same indices. ∎

### 4.3 The combinations jointly cover the upper triangle

*Theorem*

The number of distinct indices (output-matrix elements) that are jointly written to by all combinations equals the number of indices in the upper triangle of the result matrix.

*Proof*

Different combinations write to different indices, so the total number of distinct indices that all the combinations jointly write to is the sum of $\eta(\Delta r, \Delta c)$ over all combinations:

$$\eta = \sum_{\Delta r=0}^{P-1} \sum_{\Delta c=0}^{Q-1} \eta(\Delta r, \Delta c) + \sum_{\Delta r=-P+1}^{-1} \sum_{\Delta c=1}^{Q-1} \eta(\Delta r, \Delta c) \quad . \tag{4.18}$$

$$\eta_1 = \sum_{\Delta r=0}^{P-1} \sum_{\Delta c=0}^{Q-1} \eta(\Delta r, \Delta c) = \sum_{\Delta r=0}^{P-1} \sum_{\Delta c=0}^{Q-1} (P - |\Delta r|) \cdot (Q - |\Delta c|) \tag{4.19}$$

By separating the variables and then summing the series, it is possible to reduce (4.19) to:

$$\eta_1 = \frac{P}{2}(P+1)\frac{Q}{2}(Q+1) \tag{4.20}$$

Similarly,

$$\eta_2 = \frac{P}{2}(P-1)\frac{Q}{2}(Q-1) \tag{4.21}$$

Adding the two expressions from (4.20) and (4.21) gives:

$$\eta = \eta_1 + \eta_2 = \frac{PQ}{2}(PQ+1). \tag{4.22}$$

The total number of elements in the upper triangle (including the main diagonal) is:

$$Elements = \sum_{r=1}^{PQ}(P \cdot Q - (r-1)). \tag{4.23}$$

Summing this equation gives the same result as in (4.22). ∎

## 5. RESULTS

In this section, we compare implementation results of the new parallel algorithm with those of the baseline algorithm. The new algorithm was implemented twice: on a single-core system and on a shared-memory many-core system. The parallel implementation targeted the HyperCore[Plurality 2007; Plurality 2009] CREW shared-memory many-core system. In the absence of a chip at this time, Plurality's cycle-accurate simulator was used. It simulates memory accesses, memory stacks, scheduling, instruction decoding and more.

We present the results, first for a shared-memory many-core system, Plurality HyperCore and then for the x86.

The issues that we set out to assess on the two platforms are very different: on Hypercore, the issue is speedup vs. the number of cores - scaling. On the x86, in contrast we merely wanted to assess the "side benefits" of our approach, which is unintuitive even for a sequential implementation, on the performance of such an implementation relative to a naïve implementation of the algorithm. Comparison with other covariance algorithms that compute a different estimated covariance matrix is not one of our goals.
.

### 5.1 Implementation on Plurality's Many-Core System

The algorithm was implemented for the Plurality HyperCore architecture [Plurality 2007; Plurality 2009]. HyperCore features tens to hundreds of compute cores, connected to an even larger number of memory banks that jointly comprise the shared cache. The connection is via a high speed, low latency combinational interconnect. With this, one enjoys the benefits of a uniform memory architecture without the communication bottleneck of a shared bus. Also, memory coherence comes free, as there is no private memory. The memory hierarchy also includes off-chip (shared) memory. Finally, the programming model is a set of sequential "tasks'" along with a set of precedence relations among them, and these are enforced by a very high throughput, low latency hardware synchronizer/scheduler that dispatches work to the cores.

It is Plurality's goal to make this system a low power system. While exact numbers cannot be given as this platform has not been fully synthesized at the date of submission, the numbers suggest ~4 $Watts$ for 64 OpenSPARC cores at 500$MHz$ with 40$nm$ CMOS technology. Their next generation system will include 128 cores. .As such we evaluated our algorithm on Plurality's cycle-accurate simulator for various numbers of cores. This provided insights regarding load balancing and speedup trends. We note that simulator has several limitations that include a maximal cycle count. Further the simulation only considers the shared-cache without considering the DRAM. These limit the possible sizes of input matrices that can be tested on this simulator, however, the insights attained are interesting. We were able to confirm these speedup and scheduling results using simple offline scheduling techniques as we had the sequential execution times for each combination of our new algorithm on the Hypercore.
(b)
Fig. 4 depicts the speedup of the algorithm as a function of the number of cores. The comparison was carried out for two input matrix with different windows sizes:
(b)
Fig. 4 (a) presents results for input matrix of size N = M = 20 and window size P = Q = 8. For this parameter setting, there are a total of 113 combinations. (b)
Fig. 4 (b) presents results for input matrix of size N = M = 32 and window size P = Q = 13. For this parameter setting, there are a total of 313 combinations.

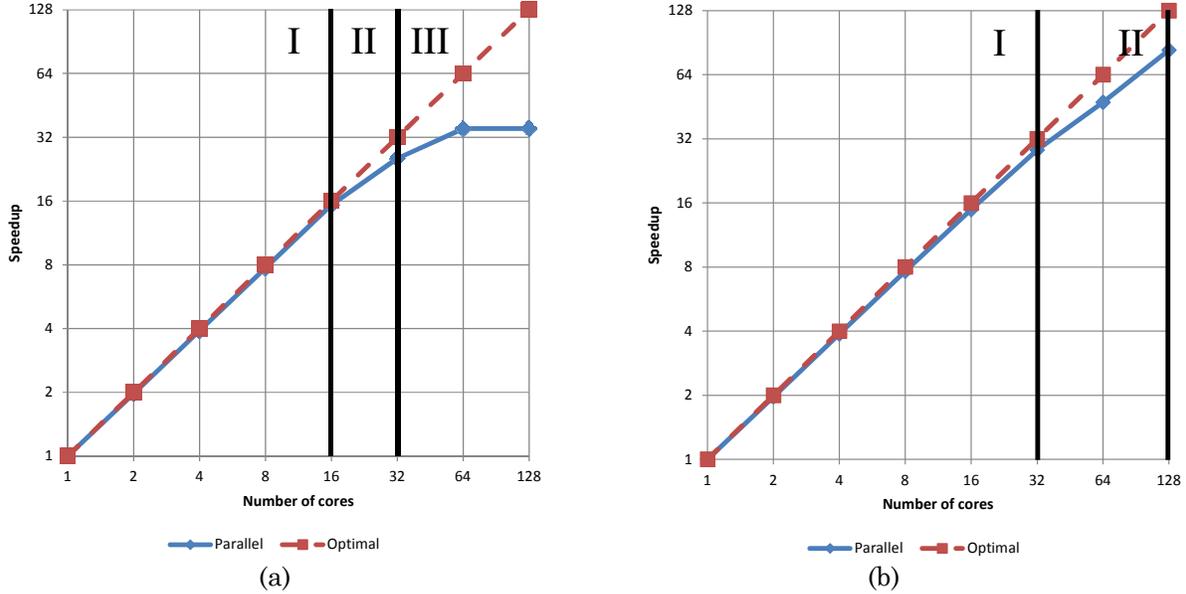

Fig. 4. Parallel execution of the our algorithm with various number of cores. (a) Input matrix: 20 × 20; window size: 8 × 8. (b) Input matrix: 32 × 32; window size:13× 13.

It is readily evident that "perfect" speedup is achieved up to some 16 cores for the smaller window size and up to 32 cores for the bigger window size. Since, as is evident from the algorithm itself, the entire algorithm has been parallelized, non-ideal speedup can only stem from memory-access problems or from imperfect load balancing among the cores. As for load balancing, since a new task (combination) is assigned to a core as soon as it becomes free, imperfect load balancing takes place when the remaining number of combinations is smaller than the number of cores or due to the different sizes of the combinations.

In the graph, three different ranges can be discerned:

Near-Optimal – In this range, the number of cores is smaller by an order of magnitude than the number of combinations. Consequently, the fraction of total execution time during which some of the cores may have no work to do is negligible relative to the total execution time. This is marker as I in (b)
Fig. 4.

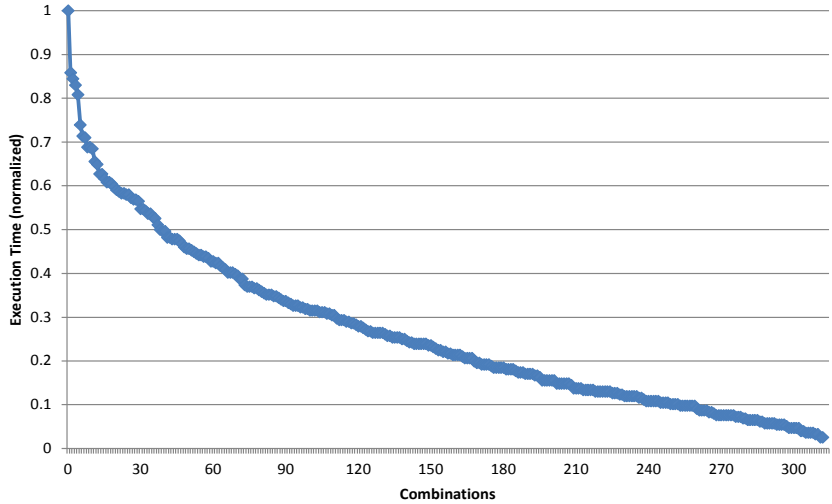

Fig. 5 – Execution times of the combinations sorted from the longest to the shortest. The executions have been normalized based on the longest combination.

Sub-Linear – In this range, most cores compute multiple combinations, but the "tail", namely completion of a "long" combination by a core that received it at a late stage, is more significant. This is marked as II in (b) Fig. 4. Note that for the bigger window size, the sub-linear speedups only start for 32 cores vs. 16 cores for the smaller window. This is due to the fact that there are more combinations that need to be computed and that the work needed by each combination is bigger. As such there is more work to go around and there is better workload balance.

Starvation – As the number of cores approaches the number of combinations (and obviously beyond it), additional cores are unlikely to contribute. This effect becomes prominent even when there are twice as many combinations as cores due to the unequal length (compute time) of combinations. This is because a core that received a short combination has time to execute another one while a core that received a long one is still processing its first combination. The addition of another core does not help, as the former core would be idle given the parallel granularity is at the combination level and that cores do not share work. This is marked as III in (b) Fig. 4 (a). Note that for the larger window this range is not denoted as the workload is balanced up to 128 cores. This would not be the case for 256 cores or 512 cores.

Fig. 5 depicts the normalized execution time for each combination. For the sake of presentation, we sorted the execution times from the longest to the shorted (though shortest to longest would simply invert the graph) for the 13x13 window. As can be seen, the bigger combinations can take up to $\sim 40X$ times more than the smallest combination. This causes a significant workload imbalance when the number of cores is greater than the number of combinations.

To overcome the starvation problem, we suggest computing several estimated covariance matrices concurrently, as is required anyhow in many signal processing applications, (E.g., [DeGraaf 1998; Yadin, Olmar, Oron and Nathansohn 2008]). This increases the workload and allows for better distribution. By using this approach, it is possible to achieve the linear speedup as desired. The ability to do so effectively also proves that there is no memory-access bottleneck. (All this has been confirmed by detailed simulations.) We note that when we added a second covariance matrix to the workload, meaning that $313 \cdot 2$ combinations were computed, the parallelism increased significantly and was nearly linear, thus starvation was avoided. For the sake of brevity, we do not present this graph.

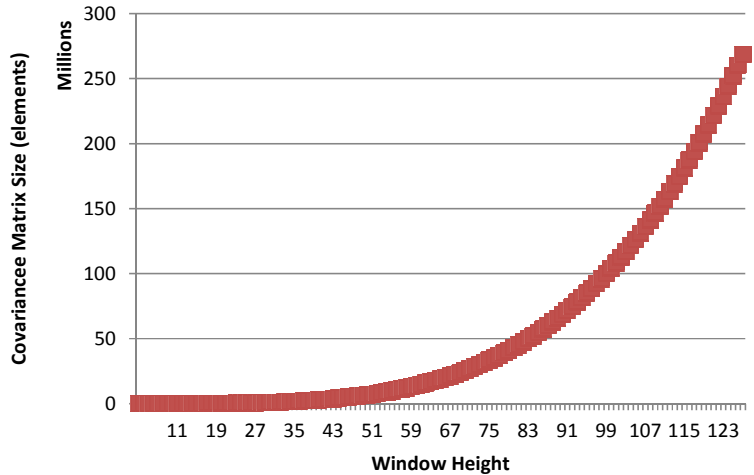

Fig. 6 – Size of the estimate covariance matrix as a function of the window height assuming a square moving window. Note that the y-axis is in millions of elements.

To the best of the authors' knowledge, there are no additional parallel algorithms for computing the estimate covariance matrix using the covariance method that do not have additional memory overhead and do not require atomic instructions. Which are not supported on the Hypercore. Therefore, no additional comparisons are presented for this platform.

**5.2 Implementation on x86 Architecture**

Our main goal in this work was effective parallelization. However, parts of our approach, such as refraining from repeating multiplications, are also applicable to serial execution. We now compare the baseline (naïve) algorithm with the new algorithm on the x86 system. The computational complexity of the two algorithms was mentioned earlier, and details are provided in Section 4.1.

In this section, actual measurements are provided. The system used is an Intel I7 quad core running at 3.4 GHz. The system has $8MB$ cache and $12GB$ of memory. The system supports Hyper-Threading, however, we did not use this feature and limited ourselves to the four physical cores.

We show results for input sizes of $32X32$ and $64x64$. While, these might seem like small input sizes, these are actual sizes of the input matrix.[DeGraaf 1998; Yadin, Olmar, Oron and Nathansohn 2008]. Increasing the input size beyond that raises computational challenges and storage challenges. In Fig. 6 the storage requirements for the estimated covariance matrix are shown as a function of the window height assuming a square window. Given the quadratic rate in which the window grows, larger input sizes are not practical for some systems. We also note that many image processing algorithms outside the scope of this work uses square image blocks of size $8x8, 16x16$, and $32x32$.

The significant storage requirements greatly limit the scalability of the most intuitive parallel approach which suggests that each thread uses a local copy of the estimated covariance matrix where the results are summed up in the end. For this approach, even when the number of threads is small is and it is possible to maintain all these copies in the memory, the maximal parallel speedup is limited by the number of threads used (though optimal performance is based on system parameters such as bandwidth, shared-memory hierarchy and such). An additional approach to overcoming the scaling issues is to maintain a single copy of the estimated covariance matrix and have the threads use synchronization methods, such as atomics and locks, to update the values. While this approach is scalable in the storage requirements, increasing the number of threads also increases the likelihood that multiple threads will try to update the same element in the estimated covariance matrix. In addition to this performance is reduced due to the overhead of the synchronization. As such for a 4-core, the maximal speedup of these approaches would also be $4X$, respectively, which is even smaller than the speedup of our algorithm without parallelism.

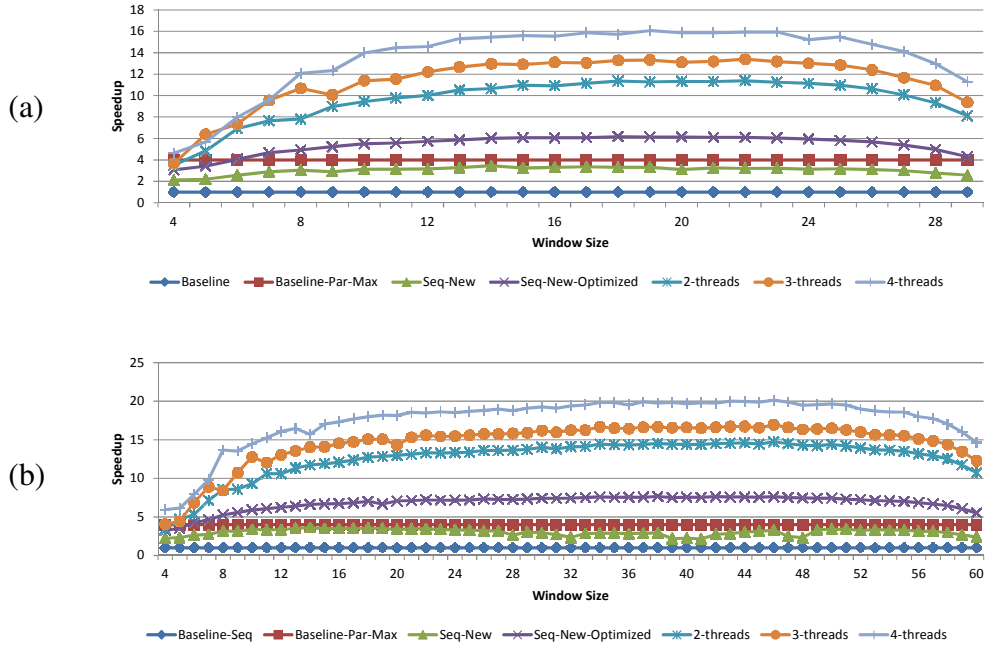

Fig. 7 – Speedup of the new algorithm as a function of the window size. (a) Input matrix: $32 \times 32$ .(b) Input matrix: $64 \times 64$.

Due to the limitations of these parallel approaches, there is no point comparing to them, especially as our new approach is significantly faster even for a single core and has all the benefits of no synchronization overhead, highly scalable, and no memory overhead. Regardless, we do present the maximal speedup that can be attained using the methods which is dependent on the number of cores available.

The main advantage of our algorithm over the naïve algorithm (and all parallel approaches based on the naïve algorithm) is the avoidance of duplicate multiplications. Duplication of additions still occurs and there are roughly equal numbers of additions and multiplications, yet the speedup is greater than 2X because multiplication takes more time than additions. In this context, we note that modern x86 systems support MAC, Multiply and Accumulate, instructions that can do a floating point multiplication and add it in a single cycle. While our implementations, in the C language, did not use these instructions directly, we did turn on the appropriate compiler flags that should be able to detect the cases and translate the C code to the appropriate assembly code that does make use of these instruction. We note that even with the use of MAC instructions the new algorithm will perform better due to a reduction in the number of instructions that need to be computed.

Fig. 7 depicts the measured speedup of our algorithm relative to the naïve algorithm, denoted as Baseline, versus window size. A direct implementation of our algorithm yields surprisingly low, speedup (Seq-New), especially for with larger window sizes. A deeper comparison of our algorithm with the baseline algorithm, however, reveals that the savings in multiplications came at the price of a loss of spatial locality of memory access. (We moreover conjecture that the dips in the curve are related to the relative sizes of the window and the cache line.)

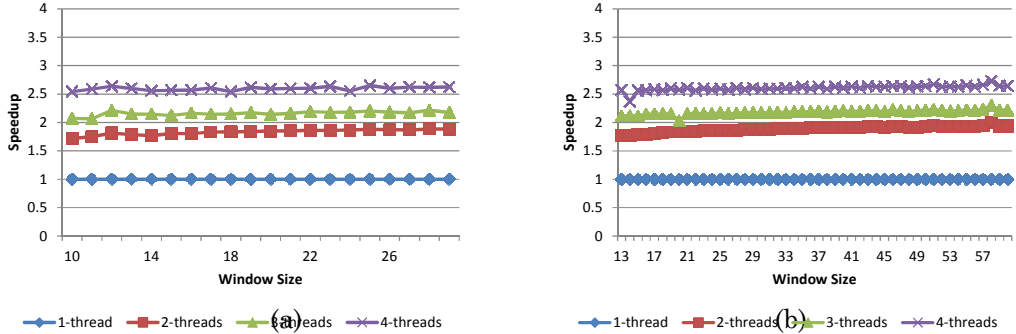

Fig. 8 – Strong scale of the new algorithm as a function of the window size. (a) Input matrix: $32 \times 32$ .(b) Input matrix: $64 \times 64$.

For our algorithm to also exhibit spatial locality in memory access (not the case when accessing diagonal segments), we allocated a serial array for each combination, and used these arrays during the computation. Once the construction of the combinations was completed, the arrays were written back to the output matrix, each array to its respective (partial) diagonal. The result (Seq-New-Optimized) has an increase in speedup in the range $6.0X - 7.2X$, a very substantial improvement over the direct implementation and a dramatic advantage over the naïve algorithm. We further note, that the sequential algorithm is faster than the maximal speedup that can be attained by parallelizing the naïve approach without the reduction in the number of multiplications. In Fig. 7 we placed a constant curve marking this maximal speedup which is $4X$ for a quad-core.

The speedups for our parallel implementation are marked as 'p'-threads, where 'p' refers to the number of threads being used: 2, 3, or 4. For the smaller input size, Fig. 7 (a), the parallel algorithm achieves a maximal speedup of $16X$ using 4 threads. For the bigger input size, Fig. 7**Error! Reference source not found.** (b), the parallel algorithm achieves a maximal speedup of $20X$ using 4 threads. The bell like shapes of the speedup curves denoting the parallel implementation are explained by the fact that as the window size increases, there is more work that needs to be done in-order to compute the estimated covariance matrix. As the window size increases and is nearer in its size to the input size, it can also be moved fewer times in the input matrix and as such our new algorithm does not significantly reduce the number of computations as each combination has little work to do and the number of multiplications per combination increases on average.

In Fig. 8, we show the strong scaling speedups of our new approach. The scaling is similar for the both the smaller input size and the larger input size.

**Remark.** HyperCore's large L1 cache (1-4MB) suffices to contain all the data for substantially larger parameter values than the L1 cache of an x86, rendering the spatial locality unimportant in many practical cases. For larger values, the same remedy can be used on the Hypercore as was done for the x86.

## 6. CONCLUSIONS

This paper presented a novel approach for parallelizing the computation of the estimated covariance matrix, an important building block for digital signal processing. Using critical insights pertaining to the relationship between input-matrix pair-wise products and the output-matrix entries to which they contribute, efficient parallelization was made possible: no multiplications are repeated, yet no coordination is required among cores and the memory footprint is very small. In so doing, we took advantage of a unique shared-memory architecture that naturally supports an otherwise awkward partitioning of memory among cores.

Experimental results show near-perfect speedup on 16 cores for a single matrix, and perfect linear speedup on as many as 128 cores when several matrices are computed concurrently. Finally, some of the insights and corresponding approaches are relevant even to single-core implementations and execution. For a quad-core x86 system a 20$X$ speedup over the baseline algorithm is achieved.

While the paper focused on a particular computation, the insights and approaches are likely to be broadly applicable.

## 7. ACKNOWLEDGEMENT

The authors are grateful to Oz Shmueli of the Parallel Systems Lab in the Electrical Engineering department for his assistance. We would also like to thank Boaz Porat for sharing with us his knowledge in signal processing.